\documentclass[runningheads]{llncs}

\usepackage{eccv}

\usepackage{eccvabbrv}
\usepackage{wrapfig}
\usepackage{cite}
\usepackage{graphicx}
\usepackage{booktabs}
\definecolor{cvprblue}{rgb}{0.21,0.49,0.74}
\usepackage{colortbl}
\usepackage{bm}
\definecolor{tablegray}{gray}{.9}
\usepackage{threeparttable}
\usepackage{longtable}
\usepackage{multirow}
\newcommand{\bs}{\boldsymbol}
\usepackage[accsupp]{axessibility}  

\usepackage{hyperref}
\usepackage{orcidlink}

\begin{document}
\begin{sloppypar}
\title{Learned HDR Image Compression for Perceptually Optimal Storage and Display} 

\titlerunning{Learned HDR Image Compression}

\author{Peibei Cao\inst{1}\orcidlink{0000-0001-7463-0409} \and
Haoyu Chen\inst{1}\orcidlink{0000-0001-8093-9648} \and
Jingzhe Ma\inst{2}\and
Yu-Chieh Yuan\inst{2} \and
Zhiyong Xie\inst{2} \and
Xin Xie\inst{2} \and
Haiqing Bai\inst{2} \and
Kede Ma\inst{1,3}\thanks{Corresponding author.}\orcidlink{0000-0001-8608-1128}}

\authorrunning{Cao~\etal}

\institute{Department of Computer Science,  City University of Hong Kong\and
Xellar Biosystems\and
Shenzhen Research Institute,  City University of Hong Kong\\
\email{\{peibeicao2-c,haoychen3-c\}@my.cityu.edu.hk\\
\{jma,yyuan,zxie,xxie,hbai\}@xellarbio.com \\
kede.ma@cityu.edu.hk}}
\maketitle

\begin{abstract}
High dynamic range (HDR) capture and display have seen significant growth in popularity driven by the advancements in technology and increasing consumer demand for superior image quality. As a result, HDR image compression is crucial to fully realize the benefits of HDR imaging without suffering from large file sizes and inefficient data handling. Conventionally, this is achieved by introducing a residual/gain map as additional metadata to bridge the gap between HDR and low dynamic range (LDR) images, making the former compatible with LDR image codecs but offering suboptimal rate-distortion performance. In this work, we initiate efforts towards end-to-end optimized HDR image compression for perceptually optimal storage and display.
Specifically, we learn to compress an HDR image into two bitstreams: one for generating an LDR image to ensure compatibility with legacy LDR displays, and another as side information to aid HDR image reconstruction from the output LDR image.
To measure the perceptual quality of output HDR and LDR  images, we use two recently proposed image distortion metrics, both validated against human perceptual data of image quality and with reference to the uncompressed HDR image. Through end-to-end optimization for rate-distortion performance, our method dramatically improves HDR and LDR image quality at all bit rates. The code is available at \url{https://github.com/cpb68/EPIC-HDR/}.
\end{abstract}

\setcounter{footnote}{0}

\section{Introduction}
\label{sec:intro}
High dynamic range (HDR) images reproduce natural scenes with a significantly wider range of luminance levels compared to low dynamic range (LDR) images.
In the past decades, HDR imaging techniques and display devices have made remarkable progress, in response to the growing need for HDR support in diverse domains, including photography and videography, virtual reality, medical imaging, autonomous driving, video surveillance, and gaming.
However, the significantly increased bit rates pose considerable challenges for HDR data storage and transmission, thereby hindering its widespread adoption, especially in real-time streaming-based HDR applications.

The most straightforward and convenient solution involves expanding the capability of existing image/video compression standards (such as JPEG, MPEG-4, H.264, and HEVC) to manage HDR data~\cite{mantiuk2004perception, garbas2011temporally, miller2013perceptual, mukherjee2018uniform,ward2006jpeg, Boschetti10flexible, mai2011optimizing, artusi2016jpeg, zaid2017hdr}. A key step in this approach is the two-layer decomposition. A base layer is constructed using manually designed global tone mapping operators (TMOs), readily displayable on LDR monitors. An extension layer, which is necessary for HDR image reconstruction as side information, is stored as either an additive residual map~\cite{artusi2016jpeg} or a multiplicative gain map~\cite{ward2006jpeg}. Both layers are compressed using existing LDR image/video compression standards, and the compressed extension layer can further be downsampled to reduce its size. The two-layer decomposition approach primarily aims to maintain backward compatibility, and often results in suboptimal rate-distortion performance. For instance, the optimal format and coding strategy for the side information that bridges the gap between HDR and LDR images have not been thoroughly investigated. Additionally, while an LDR image is generated as an intermediate result, its perceptual quality is not optimized, as it is not part of the design goal.
 
Another active area of research focuses on compressing only the dynamic range of HDR images, making them displayable on standard LDR monitors by designing HDR image TMOs~\cite{drago2003adaptive,reinhard2002photographic,mai2011optimizing}. Equipped with inverse TMOs (iTMOs)~\cite{banterle2006inverse}, it is feasible to reconstruct HDR images. However, this approach prioritizes the visual quality of tone-mapped images without pursuing the optimal rate-distortion compromise. Again, existing LDR image/video codecs may come to the rescue but at the cost of visual quality degradation. Additionally, the cascaded and separate design and optimization of TMOs and iTMOs may sacrifice the visual quality of reconstructed HDR images.

To address some of the above-mentioned drawbacks, deep learning has been explored for HDR image compression. Cao~\etal~\cite{cao2022oodhdr} and Han~\etal~\cite{han2020hdr} trained deep neural networks (DNNs)  to compress the based layer and the extension layer, respectively. Existing HDR image compression systems with learnable components exhibit several notable limitations: they are not 1) end-to-end optimized by 2) perceptually aligned HDR and LDR image distortion metrics for 3) both storage and display purposes. 

\begin{figure}[!t]
  \centering
    \subfloat[Traditional HDR image compression systems based on the two-layer decomposition]{\includegraphics[width=0.99\linewidth]{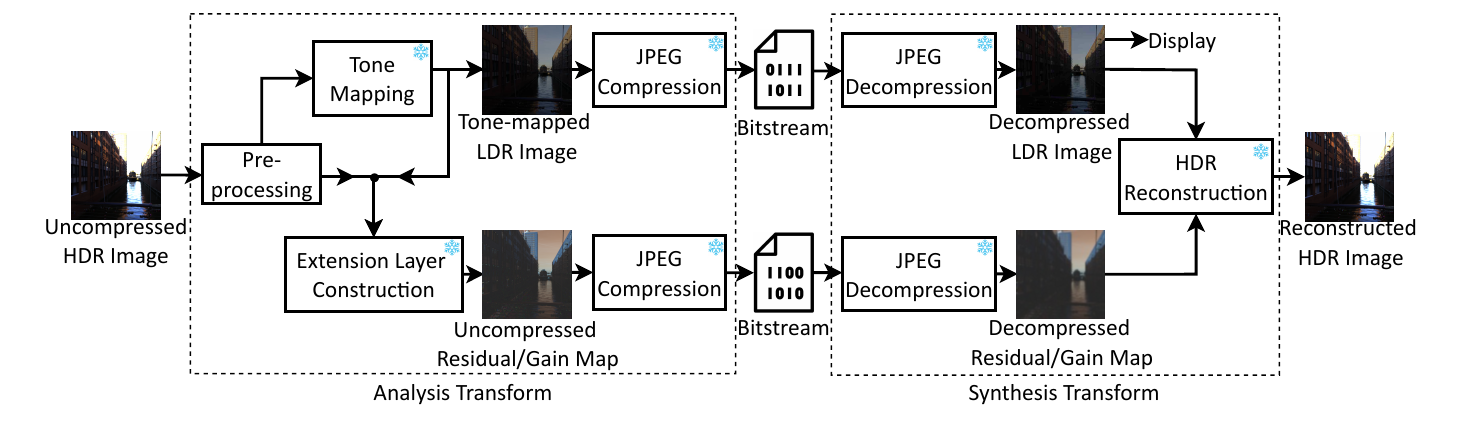}}
    \vfill
    \subfloat[Proposed EPIC-HDR for perceptually optimal storage and display]{\includegraphics[width=\linewidth]{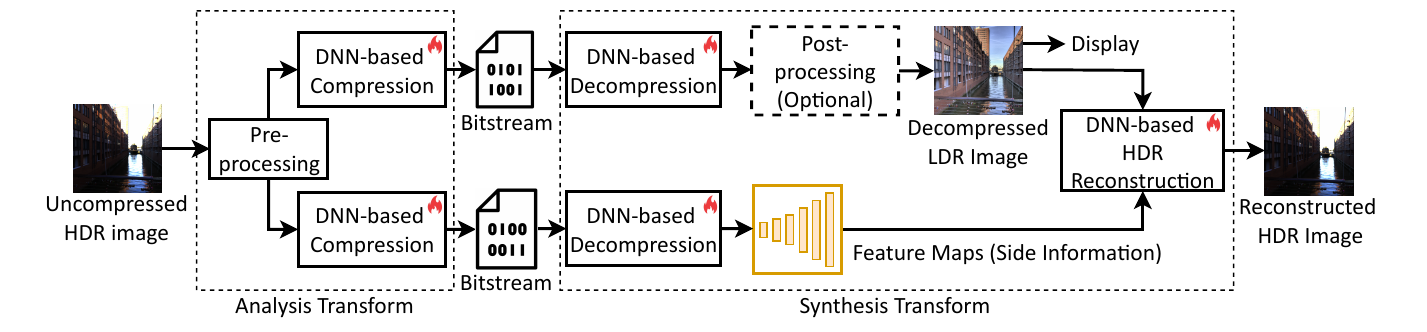}}
  \caption{Comparison between (a) the traditional manually designed and (b) the proposed learned HDR image compression systems.}
  \label{fig:diag}
\end{figure}

In this work, we aim ambitiously for learned HDR image compression for perceptually optimal storage and display. Following the transform coding paradigm, we design a pair of analysis and synthesis networks to extract  LDR latent codes from the input HDR image and subsequently generate an LDR image, respectively. The synthesis network is conditioned on the maximum scene luminance, similar to physics-based TMOs~\cite{reinhard2005dynamic,zhang2020retina,cao2022perceptually}.
Simultaneously, we construct another pair of analysis and synthesis networks to extract HDR latent codes from the same input HDR image and generate a set of feature maps as side information to aid HDR image reconstruction. We train one more reconstruction network that takes the output LDR image and reconstructs the HDR image, conditioned on the synthesized feature maps. Both the HDR and LDR latent codes are quantized before being fed into the synthesis networks. Their actual rates are estimated by an entropy model leveraging joint autoregressive and hierarchical priors~\cite{minnen2018joint},
and then saved into bitstreams using arithmetic coding.

To assess the perceptual quality of the output HDR images, we employ a newly proposed HDR image distortion metric~\cite{Cao2024perceptual} based on a simple inverse display model~\cite{mantiuk2009visualizing}, which allows for detailed comparison by zooming into local luminance ranges. To assess the perceptual quality of the output LDR images,
we adopt the normalized Laplacian pyramid distance (NLPD)~\cite{laparra2017perceptually}, which is known for its effectiveness in cross-dynamic-range image quality assessment and has proven effective in optimizing TMOs~\cite{laparra2017perceptually,cao2022perceptually}.

Our contributions can be summarized in three key points.
\begin{itemize}
    \item We develop a learned 
 HDR image compression system, which we name End-to-end and Perceptually optimized Image Coder for HDR support (EPIC-HDR), and an automated extension, for both storage and display purposes. EPIC-HDR can be viewed as a learnable generalization of traditional HDR image compression systems based on layer decomposition (see Fig.~\ref{fig:diag}).
    \item We adopt two perceptually aligned image distortion metrics to ensure that the output HDR and LDR images are optimal with respect to human perception of image quality.
    \item We conduct extensive experiments to show the effectiveness of EPIC-HDR, highlighting its superior rate-distortion performance and dramatic improvements in HDR and LDR image quality at various bit rates.
\end{itemize}

\section{Related Work}
In this section, we review techniques for HDR image compression and learned LDR image compression.

\subsection{HDR Image Compression}

\noindent\textbf{HDR Image Compression for Storage.}
One popular line of research is to enhance standardized LDR image/video compression systems with HDR capability. Examples include JPEG2000~\cite{xu05high}, MPEG-4~\cite{mantiuk2004perception}, and H.264~\cite{garbas2011temporally}. The latest video compression standard, H.266/VVC, offers direct HDR support.
Guleryuz~\etal~\cite{guleryuz2022sandwiched}
proposed sandwiched image compression, which involves training a pre-processor to encode HDR images into compatible formats with LDR image/video codecs, and a post-processor for HDR image reconstruction. A differentiable proxy for the standard in-between LDR image/video codec shall be used to enable end-to-end optimization.

\noindent\textbf{HDR Image Compression for Display.} Another line of research only compresses the dynamic range of HDR images to ensure compatibility with existing LDR displays by HDR image TMOs. Global TMOs~\cite{tumblin1993tone, drago2003adaptive, reinhard2005dynamic, kim2008consistent} apply the same parametric functions to all pixels in an HDR image, whose parameters are determined by global image statistics (\eg, the log-average luminance~\cite{reinhard2002photographic}). In contrast, local TMOs~\cite{durand2002fast, paris2011local, liang2018hybrid} maintain relative contrast between neighboring pixels, which is more perceptible to the human eye. A common design strategy for local TMOs is two-layer decomposition~\cite{durand2002fast}, rooted in the retinex theory~\cite{land1971lightness}. Tone compression is applied to the base layer, while detail reproduction is carried out in the detail layer. Recently, DNN-based TMOs~\cite{rana2020deep,yang2021deep} have emerged by addressing the challenge of lacking paired HDR-LDR images for supervised training. TMOs can be combined with iTMOs~\cite{banterle2006inverse} (also known as single-image HDR reconstruction methods) to recover HDR images from tone-mapped LDR images. However, this cascaded design may introduce unwanted distortions regarding color and detail reproduction in the reconstructed HDR images.

\noindent\textbf{HDR Image Compression for  Storage and Display.}
To supply displayable LDR images during HDR image compression, a similar two-layer decomposition was applied~\cite{ward2006jpeg, Boschetti10flexible, mai2011optimizing,artusi2016jpeg, zaid2017hdr}. The base layer, typically obtained by a global TMO, is compressed by existing image/video codecs for backward compatibility, such as  JPEG in~\cite{ward2006jpeg,artusi2016jpeg,zaid2017hdr}, JPEG2000 in~\cite{Boschetti10flexible}, and H.264 in~\cite{LEE2012Rate}. Common formats for representing the extension layer include the residual map~\cite{artusi2016jpeg} (\ie, the difference between the uncompressed and reconstructed HDR images from the base layer) and the ratio map (\ie, the ratio between the uncompressed HDR image and the base LDR image).
These methods exhibit suboptimal rate-distortion performance due to their excessive reliance on existing codecs. 
Recently, Cao~\etal~\cite{cao2022oodhdr} and Han~\etal~\cite{han2020hdr} enhanced rate-distortion performance by employing DNNs to compress the base and extension layers, respectively.  
The proposed EPIC-HDR can be seen as a learnable generalization of the layer decomposition approach, end-to-end learned for perceptually optimal storage and display.

\subsection{Learned LDR Image Compression}
Learned LDR image compression systems also follow the classic transform coding paradigm, which includes analysis transform, quantization, entropy coding, dequantization, and synthesis transform. 

\noindent\textbf{Analysis/Synthesis Transform.} Convolutional DNNs are the most popular architecture for implementing analysis and synthesis transforms. Ball\'{e}~\etal~\cite{balle2016end} introduced generalized divisive normalization, a bio-inspired nonlinearity, to enhance the expressiveness of shallow DNNs. Cheng~\etal~\cite{cheng2020learned} incorporated the attention mechanism with significantly improved rate-distortion performance. In addition to DNNs, Transformers~\cite{zhu2022transformer}, normalizing flows~\cite{zhang2021iflow}, and diffusion models~\cite{yang2024lossy} have also been explored. The analysis and synthesis transforms of EPIC-HDR are largely based on the network architectures in~\cite{cheng2020learned}.

\noindent\textbf{Quantizer.} Differentiable approximation of quantization is a core step in learned image compression. Theis~\etal~\cite{theis2017lossy} used the straight-through estimator (\ie, the identity function) as a rudimentary proxy for quantization. Ball\'{e}~\etal~\cite{balle2017endtoend} suggested to add uniform noise from a signal processing perspective, while Agustsson~\etal~\cite{agustsson2017soft} explored soft-to-hard annealing from a numerical optimization perspective using a parametric sigmoid function, which was later improved in~\cite{guo2021soft}. In our work, we use the uniform noise addition method~\cite{balle2017endtoend} as a continuous relaxation of discrete quantization.

\noindent\textbf{Rate Function.} One natural way to bound the rate is by using the number of latent codes~\cite{toderici2016full,li2020learning}. A more mathematically sound approach approximates the rate as the entropy of the quantized codes, which are assumed to follow some parametric probability distributions, including channel-wise piecewise linear functions~\cite{balle2016end}, and code-wise Gaussians~\cite{balle2018variational} and mixtures of Gaussians~\cite{li2020efficient}. The parameters of these distributions are estimated by DNNs~\cite{balle2018variational,minnen2018joint}. In our work, we adopt code-wise Gaussians~\cite{balle2018variational} for rate estimation.

\noindent\textbf{Distortion Function.} Conventional distortion measures, such as mean squared error (MSE) and structural similarity (SSIM) index, are frequently used. Furthermore, task-oriented losses, such as adversarial loss~\cite{agustsson2019generative} for improving perceptual image quality at low bit rates and cross-entropy loss~\cite{torfason2018towards} for boosting image recognition performance, can also be included. In our work, we employ two perceptually aligned image distortion metrics to ensure the perceptually optimal storage and display of HDR images. 

\begin{figure*}[!t]
  \centering
  \includegraphics[width=.9\linewidth]{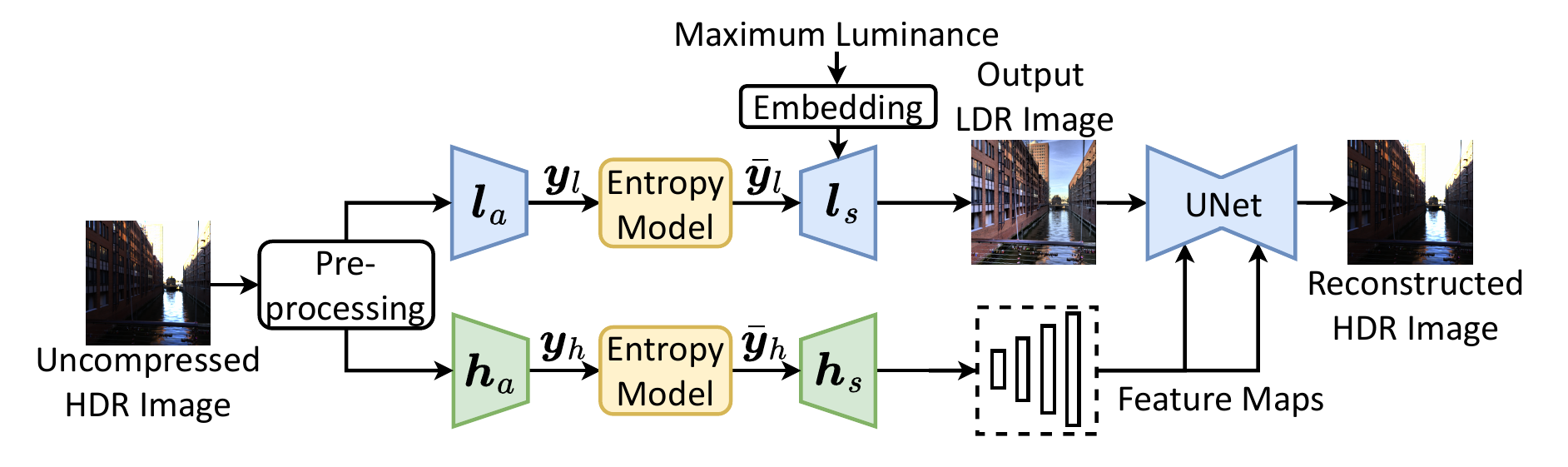}
  \caption{System diagram of EPIC-HDR. $\bs{l}_a(\cdot)$ and $\bs{h}_a(\cdot)$ are analysis networks, while $\bs{l}_s(\cdot)$ and  $\bs{h}_s(\cdot)$ are synthesis networks.}
  \label{fig:strc}
\end{figure*}

\section{Proposed Method: EPIC-HDR}
In this section, we present EPIC-HDR, our learned HDR image compression method for perceptually optimal storage and display. EPIC-HDR includes two pairs of synthesis and analysis networks and an HDR image reconstruction network. We highlight the use of two perceptually aligned image distortion metrics. Additionally, we introduce the automated extension of EPIC-HDR. The system diagram of EPIC-HDR is shown in Fig.~\ref{fig:strc}. 
\subsection{EPIC-HDR}
\noindent\textbf{Pre-processing.} For perceptual HDR image processing, it is desired to work with \textit{calibrated} HDR images that capture actual luminance values in the physical unit of $\mathrm{cd}/\mathrm{m}^2$ (nits). This is due to the highly nonlinear responses of the human eye to varying light levels~\cite{carandini2012}. 
Unfortunately, most available HDR images are uncalibrated, meaning their recorded measurements are linearly proportional to the actual luminances with an unknown scaling factor. Estimating the maximum luminance of an HDR image for the optimal rate-distortion performance during preprocessing is highly complex.
Therefore, we choose to linearly rescale the luminance values of uncalibrated HDR images to the range $[0, 1]$, and incorporate the maximum scene luminance as the conditional information into the LDR synthesis network,  allowing for flexible maximum luminance adjustment during post-processing (see Fig.~\ref{fig:diag}).

\begin{figure}[!t]
  \centering
  \includegraphics[width=0.7\linewidth]{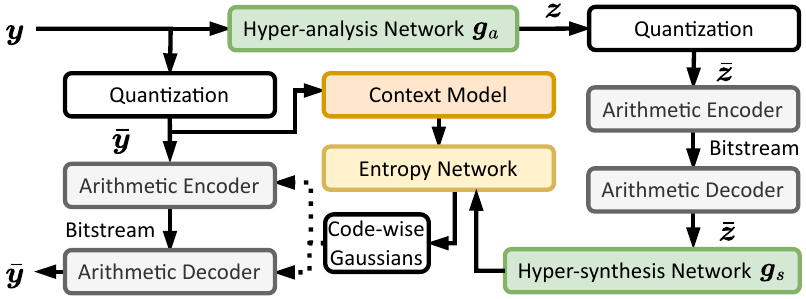}
  \caption{Computational structure of the entropy model. $\bs{g}_a$ is the hyper-analysis network and $\bs{g}_s$ is the hyper-synthesis network. The entropy network is used to estimate Gaussian parameters (\ie, the means and variances).}
  \label{fig:entr}
\end{figure}

\noindent\textbf{Analysis, Synthesis, and Reconstruction Networks.} We adopt two analysis networks, $\bs{h}_a(\cdot)$ and $\bs{l}_a(\cdot)$, to transform the pre-processed HDR image into HDR and LDR latent codes, denoted as $\bs{y}_h$ and $\bs{y}_l$, respectively. 
These latent codes are then quantized to $\bar{\bs{y}}_h$ and $\bar{\bs{y}}_l$~\cite{balle2016end, balle2018variational, minnen2018joint, cheng2020learned} using a uniform quantizer (\ie, $Q(\xi) =\lfloor \xi +0.5\rfloor$). We encode the quantized codes into bitstreams via arithmetic coding for storage and transmission. 

Furthermore, we input $\bar{\bs{y}}_l$ into the LDR synthesis network $\bs{l}_s(\cdot)$ to output an LDR image, conditioned on the embedding of the maximum scene luminance. We adopt the embedding strategy described in~\cite{ho2020denoising}, initially proposed for embedding the time step in diffusion models. We assume a fixed LDR display with the minimum and maximum luminances of $1$ and $300$ $\rm cd/m^{2}$, respectively, which are common specifications for consumer-grade displays with standard dynamic ranges. Simultaneously, we input $\bar{\bs{y}}_h$ into the HDR synthesis network $\bs{h}_s(\cdot)$ to generate a set of feature maps, which serve as side information to aid HDR image reconstruction. 
These feature maps are concatenated with the intermediate feature maps of the reconstruction network $\bs{r}(\cdot)$, which accepts the generated LDR image as input to reconstruct the HDR image.

\noindent\textbf{Entropy Model.} Fig.~\ref{fig:entr} shows the computational structure of the entropy model~\cite{cheng2020learned}, with subscripts $\{l,h\}$ omitted for clarity. The hyper-analysis network $\bs{g}_a(\cdot)$ condenses the latent codes $\bs{y}$ as the hyper-prior, denoted by $\bs{z}$. Both $\bs{y}$ and $\bs{z}$ are quantized to $\bar{\bs{y}}$ and $\bar{\bs{z}}$, respectively. Following~\cite{balle2018variational},  we model $\bar{\bs{z}}$ using a factorized probability function and $\bar{\bs{y}}$ using independent Gaussians.  The distribution parameters are estimated through an entropy network, which takes both the hyper-prior $\bar{\bs{z}}$ and the autoregressive context~\cite{minnen2018joint} of $\bar{\bs{y}}$ as input.

\subsection{Rate-Distortion Function}
\noindent\textbf{Rate Function.}
We denote the conditional probability  of $\bar{\bs{y}}$ as $p(\bar{\bs{y}} \vert \bs{g}_{s}(\bar{\bs{z}}),\bs{c}(\bar{\bs{y}}))$,
where $\bs{c}(\bar{\bs{y}})$ represents the autoregressive context. 
The probability of $\bar{\bs z}$ is estimated by the factorized probability function $p(\bar{\bs z} ; \bs{\theta})$, parameterized by $\bs{\theta}$. Consequently, the rate for the LDR image is given by
\begin{equation}
\begin{aligned}
    r_\mathrm{L}(\bar{\bs{y}}_l, \bar{\bs{z}}_l) = \mathbb{E}\left[-\log_{2} p(\bar{\bs{y}}_{l} \vert \bs{g}_{s}(\bar{\bs{z}}_{l}), \bs{c}(\bar{\bs{y}}_{l}))\right] 
    +\mathbb{E}\left[-\log_{2} p(\bar{\bs z}_l ; \bs{\theta}) \right].
\end{aligned}
\end{equation}
The rate for the HDR side information $r_\mathrm{H}(\bar{\bs{y}}_h)$ is computed similarly, where we make a direct estimation without the need for constructing the hyper-prior $\bar{\bs z}_h$. 

\noindent\textbf{HDR Image Distortion Function.}
We adopt the recently proposed HDR image distortion metric~\cite{Cao2024perceptual}, which relies on an inverse display model~\cite{mantiuk2009visualizing} to decompose the reference HDR image, denoted by  $\bs{S}$, into a stack of $K$ LDR images $\{\bs{I}^{(k)}\}_{k=1}^K$ with varying exposure values. Similarly, we decompose a test HDR image $\hat{\bs{S}}$ into $\{\hat{\bs{I}}^{(k)}\}_{k=1}^K$, and compute its perceptual distortion relative to the reference by
\begin{align}\label{eq:gw}
     d_\mathrm{H}\left(\bs{S},\hat{\bs{S}};\{{e}^{(k)}\}_{i=1}^K,\{\hat{{e}}^{(k)}\}_{i=1}^K\right)= \sum_{k=1}^{K}d\left(\bs{I}^{(k)},\hat{\bs{I}}^{(k)};{e}^{(k)},\hat{{e}}^{(k)}\right),
\end{align}
where ${e}^{(k)}$ and $\hat{{e}}^{(k)}$, for $1\le k\le K$, denote the $k$-th exposure value for the reference and test HDR images, respectively. $d(\cdot,\cdot)$ represents a mature LDR image distortion metric, which is instantiated by the locally adaptive deep image structure and texture similarity (ADISTS) index~\cite{ding2021locally} in our paper. 

When using $d_\mathrm{H}(\cdot,\cdot)$ in Eq.~\eqref{eq:gw} as the loss function, we set ${e}^{(k)} = \hat{{e}}^{(k)}$, for $1\le k\le K$. However,
when adopting $d_\mathrm{H}(\cdot,\cdot)$ as the evaluation metric, it is necessary to mitigate potential luminance shifts between the reference and test HDR images by
\begin{align}\label{eq:omeh}
d_\mathrm{H}^{\star}\left(\bs{S},\hat{\bs{S}};\{{e}^{(k)}\}_{i=1}^K\right)= \min_{\{\hat{{e}}^{(k)}\}_{k=1}^{K}}d_\mathrm{H}\left(\bs{S},\hat{\bs{S}};\{{e}^{(k)}\}_{i=1}^K,\{\hat{{e}}^{(k)}\}_{i=1}^K\right).
\end{align}
$d^\star_H(\cdot,\cdot)$ has been demonstrated effective in evaluating HDR image quality and optimizing HDR new view synthesis methods~\cite{Cao2024perceptual}.

\begin{figure*}[!t]
  \centering
  \begin{subfigure}{0.19\textwidth}
      \includegraphics[width=\textwidth]{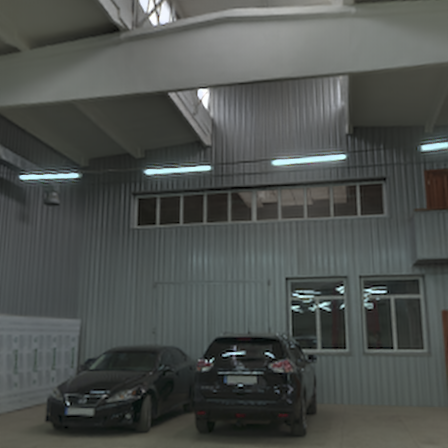}
      \caption{$10^{4}$ $\rm cd/m^{2}$}
  \end{subfigure}
  \hfill
  \hspace{-5pt}
  \begin{subfigure}{0.19\textwidth}
      \includegraphics[width=\textwidth]{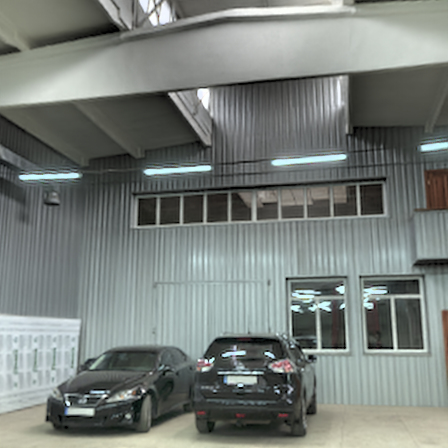}
      \caption{$10^{5}$ $\rm cd/m^{2}$}
  \end{subfigure}
  \hfill
  \hspace{-5pt}
  \begin{subfigure}{0.19\textwidth}
      \includegraphics[width=\textwidth]{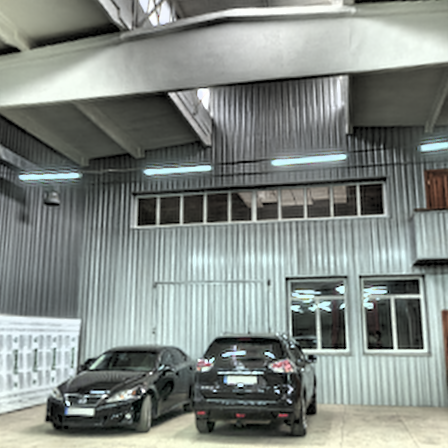}
      \caption{$10^{6}$ $\rm cd/m^{2}$}
  \end{subfigure}
  \hfill
  \hspace{-5pt}
  \begin{subfigure}{0.19\textwidth}
      \includegraphics[width=\textwidth]{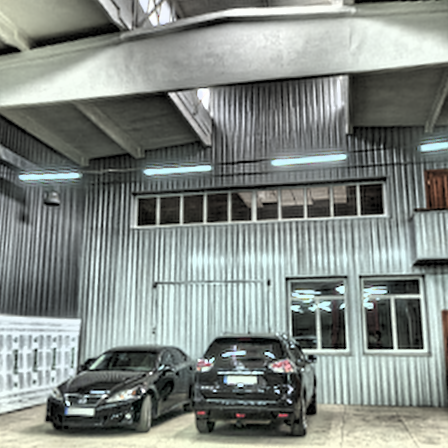}
      \caption{$10^{7}$ $\rm cd/m^{2}$}
  \end{subfigure}
  \hfill
  \hspace{-5pt}
  \begin{subfigure}{0.19\textwidth}
      \includegraphics[width=\textwidth]{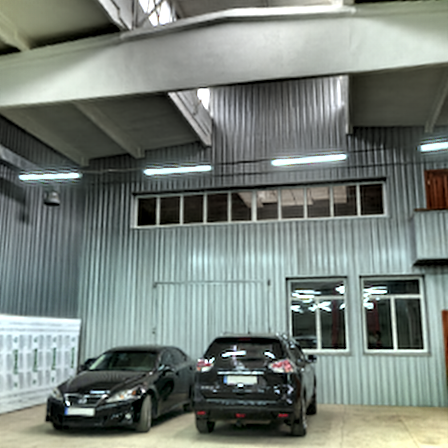}
      \caption{Fused image}
  \end{subfigure}
  
  \caption{(a)-(d) Stack of four pseudo-multi-exposure LDR images generated with different maximum luminances $\{10^{4}, 10^{5}, 10^{6}, 10^{7}\}$ $\rm cd/m^{2}$, and (e) the corresponding fused LDR image by~\cite{Li2020fast}.}
  \label{fig:extldr}
\end{figure*}

\noindent\textbf{LDR Image Distortion Function.}
We employ the NLPD metric~\cite{laparra2017perceptually} to assess LDR image distortion. NLPD quantifies the perceptual distance between an LDR image and its reference HDR image using the normalized Laplacian pyramid representations: 
\begin{align} \label{eq:nlpd}
d_\mathrm{L}(\bs{S},\hat{\bs{I}}) = \left[\frac{1}{M} \sum_{i=1}^{M}\left(\frac{1}{N^{(i)}}\sum_{j=1}^{N^{(i)}}\left\vert \bs{y}_{j}^{(i)}-\hat{\bs{y}}_{j}^{(i)}\right\vert^{\alpha}\right)^{\frac{\beta}{\alpha}}\right]^{\frac{1}{\beta}},
\end{align}
where $M$ represents the number of subbands in the pyramid, and $N^{(i)}$ indicates the number of coefficients in the $i$-th subband. $\bs{y}$ and $\hat{\bs{y}}$ correspond to the subbands from the normalized Laplacian pyramid for the HDR and LDR images, respectively. In our paper, we set $\alpha=\beta = 1$.

\noindent\textbf{Overall Rate-Distortion Function.}
Finally, the overall rate-distortion function is the weighted summation of the previously defined losses:
\begin{align}\label{eq:rd1}
    \ell =  r_\mathrm{H}(\bar{\bs{y}}_{h}) + \lambda_\mathrm{H}d_\mathrm{H}(\bs{S}, \hat{\bs{S}}) +r_\mathrm{L}(\bar{\bs{y}}_{l}, \bar{\bs{z}}_{l}) + \lambda_\mathrm{L}d_\mathrm{L}(\bs{S}, \hat{\bs{I}}),
\end{align}
where the rate-distortion trade-off is controlled by the scalar parameters $\lambda_\mathrm{H}$ and $\lambda_\mathrm{L}$, with each pair of values corresponding to a different bit rate.

\subsection{An Automated Extension of EPIC-HDR}
The current EPIC-HDR relies on an educated guess of the maximum scene luminance as the condition to generate the LDR image. To fully automate EPIC-HDR, inspired by~\cite{cao2022perceptually}, we resort to multi-exposure image fusion~\cite{mertens2007exposure} as an optional post-processing step. Specifically, the LDR synthesis network is conditioned on each of the four maximum scene luminances, $\{10^{4}, 10^{5}, 10^{6}, 10^{7}\}$ $\rm cd/m^{2}$, resulting in a stack of pseudo-multi-exposure LDR images with varying detail visibility for the input HDR image (see Fig.~\ref{fig:extldr}). These images are then combined using a top-performing off-the-shelf multi-exposure image fusion method~\cite{Li2020fast} to produce the output LDR image, and meanwhile fed into the HDR reconstruction network, along with the HDR side information $\bm h_s(\bar{\bs y}_h)$.

\begin{figure*}[t]
  \centering
    
    \subfloat[HDR-VDP-3]{\includegraphics[height=0.51\textwidth]{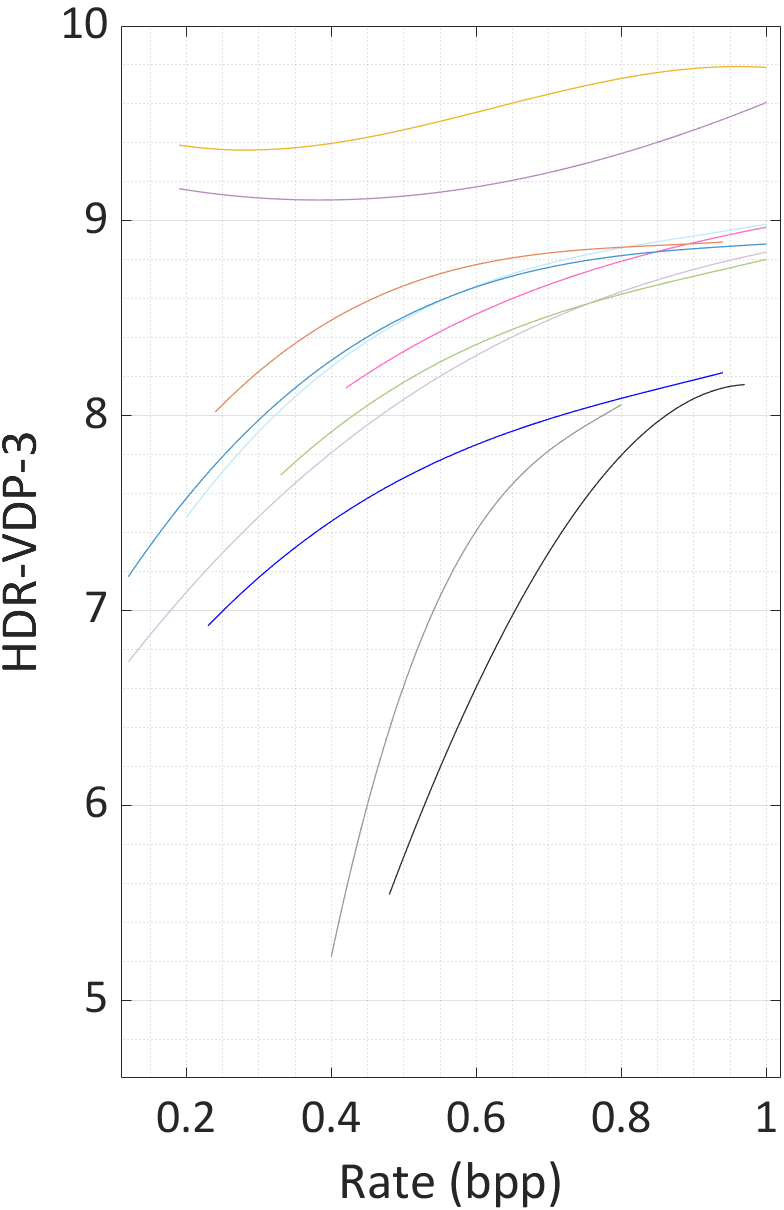}}
    \hfill
    \subfloat[PU21-CRF-PSNR]{\includegraphics[height=0.51\textwidth]{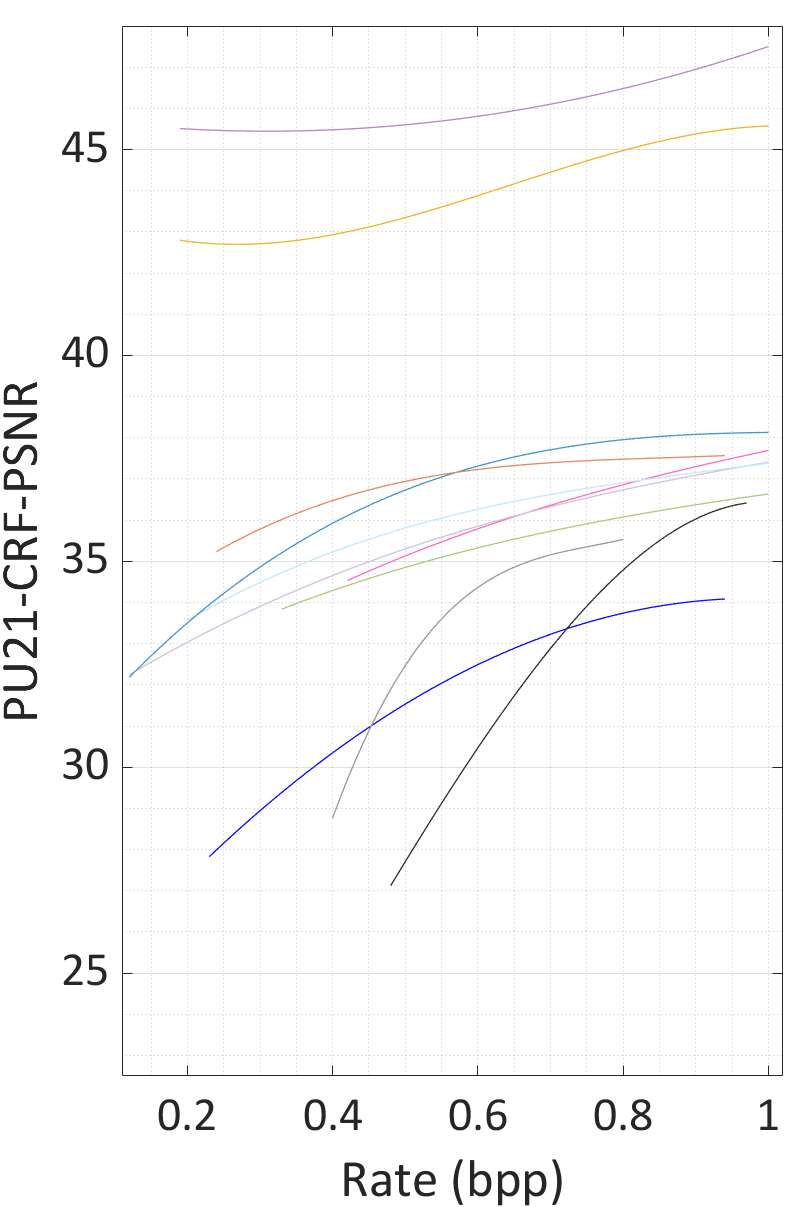}} 
    \hfill
    \subfloat[$d^\star_\mathrm{PSNR}$]{\includegraphics[height=0.51\textwidth]{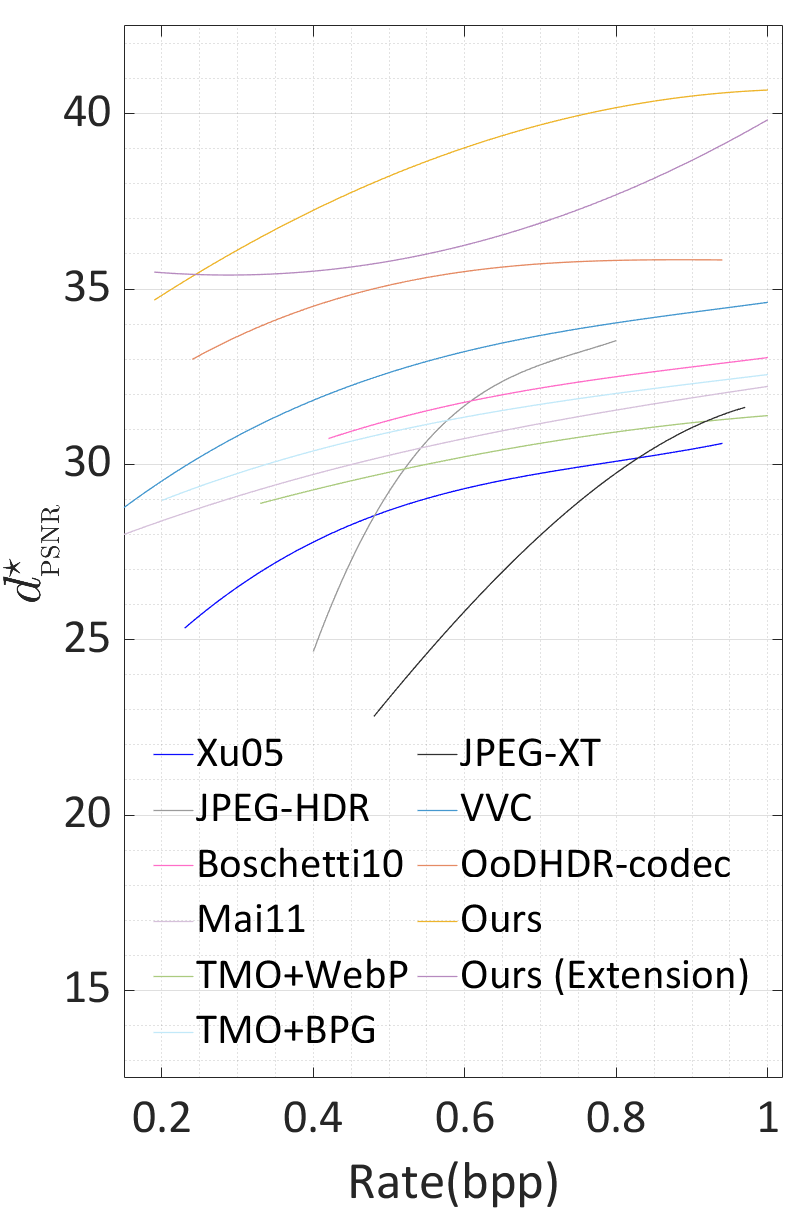}} 
    
  \caption{HDR rate-distortion performance evaluation.}
  \label{fig:hdr}
\end{figure*}

\section{Experiments}
In this section, we first describe the experimental setups, and then provide quantitative and qualitative comparison of EPIC-HDR and its extension with the state-of-the-art, followed by a series of ablation studies.
\subsection{Experimental Setups}

\noindent\textbf{Dataset.} We use panoramic HDR images from \textit{Poly Haven}~\cite{cao2022oodhdr}, each converted to ten $448\times 448$ central-perspective images with a corresponding viewing angle  of $\pi/4\times \pi/4$. In total, $3,880$ images are generated for training, and $780$ images are used for testing, ensuring content independence.

\noindent\textbf{Training and Testing.}
EPIC-HDR is optimized using Adam~\cite{kingma2014adam}, with an initial learning rate of $10^{-4}$ and a decaying factor of $10$ for every $200$ epochs. We train EPIC-HDR for a total of $600$ epochs. During training, the conditional maximum scene luminance is randomly sampled from  $\{10^{4},10^{5},10^{6},10^{7}\}$ $\rm cd/m^{2}$.
We set the LDR trade-off parameter $\lambda_{L}$  to five different values $\{20, 50, 100, 150, 200\}$, while fixing the HDR trade-off parameter $\lambda_{H}$ to $1,500$ to achieve five different bit rates.

We select nine HDR image compression methods for comparison: Xu05~\cite{xu05high}, JPEG-HDR~\cite{ward2006jpeg}, Boschetti10~\cite{Boschetti10flexible}, Mai11~\cite{mai2011optimizing}, TMO+BGP~\cite{mai2011optimizing}, TMO+WebP~\cite{mai2011optimizing}, JPEG-XT~\cite{artusi2016jpeg}, VVC~\cite{Bross2021VVC}, and OoDHDR-codec~\cite{cao2022oodhdr}. 
JPEG-HDR and JPEG-XT are HDR extensions of  JPEG, while Xu05 and Boschetti10 are HDR extensions of  JPEG2000.  
TMO+BPG and TMO+WebP are two variants of Mai11 that replace  H.264/AVC   with BPG~\cite{bpg} and WebP~\cite{webp}, respectively. OoDHDR-codec is a learned HDR image compressor. All competing methods produce LDR images, except for Xu05 and VVC.

 For HDR image quality evaluation, we employ 1) HDR-VDP-3~\cite{mantiuk2023hdr}, 2) HDR-VQM, 3) PU21-PSNR with camera response function (CRF) correction~\cite{Hanji2022},  4) PU21-SSIM with CRF correction, 5) $d^{\star}_\textrm{PSNR}$, and 6) $d^{\star}_\textrm{SSIM}$.
 For LDR image quality evaluation, we adopt 1) the tone-mapped image quality index (TMQI)~\cite{yeganeh2012objective} and 2) NLPD~\cite{laparra2017perceptually}. 

\begin{table}[!t]
  \centering
  \tabcolsep=0.15em 
    \caption{BD-Quality results, where TMO+BPG is the anchor.} \label{tab:hdr-bd}
    \resizebox{\linewidth}{!}{
    \begin{tabular}{lccccccc}
        \toprule
        \multirow{2}{*}[-3pt]{Method} &\multirow{2}{*}[-3pt]{HDR-VDP-3} &\multirow{2}{*}[-3pt]{HDR-VQM}&\multicolumn{2}{c}{With CRF correction} &\multirow{2}{*}[-3pt]{$d^{\star}_\textrm{PSNR}$}&\multirow{2}{*}[-3pt]{$d^{\star}_\textrm{SSIM}$}\\ 
        \cmidrule(lr){4-5} & & &\textrm{PU21-PSNR} & \textrm{PU21-SSIM} \\ 
        \hline
        Xu05 & -0.7798 & -0.2079 & -4.5186 & -0.0187 & -2.5083 & -0.0225\\
        JPEG-HDR & -1.3830 & -1.9401 & -2.3818 & -0.0411 & -0.1981 & -0.0544 \\
        Boschetti10 & -0.1050 & -0.1699 & -0.2982 & -0.0058 & 0.4241 & 0.0054\\
        Mai11 & -0.3800 & -0.3993 & -0.4489 & -0.0055 & -0.5948 & -0.0090\\
        TMO+WebP & -0.2851 & -0.2695 & -0.8942 & -0.0064 & -1.1213 & -0.0120\\
        JPEG-XT & -1.6587 & -2.2911 & -4.2727 & -0.0485 & -4.1449 & -0.0937\\
        VVC & 0.0208 & -1.0100 & 0.6372 & -0.0259 & 1.4146 & -0.0161\\
        OoDHDR-codec & 0.1703 & 0.1276 & 1.0331 & 0.0089 & 3.9764 & 0.0383\\
        \hline
        Ours  & \textbf{1.1202} & \textbf{1.6049} & \textbf{7.9145} & \textbf{0.0361}  & \textbf{7.6275} & \textbf{0.0612} \\
        Ours (Extension) & \textbf{0.9521} & \textbf{0.8660} & \textbf{9.7701} & \textbf{0.0283} & \textbf{5.0295} & \textbf{0.0591} \\ 
        \bottomrule
    \end{tabular}}
\end{table}

\subsection{Quantitative Comparison}
\noindent\textbf{HDR Image Comparison.}
Fig.~\ref{fig:hdr} presents the average HDR rate-distortion curves. It is clear that EPIC-HDR and its extension outperform all competing methods under all evaluation metrics and across all bit rates. 
Table~\ref{tab:hdr-bd} summarizes the Bj{\o}ntegaard-delta quality (BD-Quality) results~\cite{bjontegaard2001calculation}, where TMO+BPG is selected as the anchor. EPIC-HDR and its extension secure the top two positions by wide margins. We consider these improvements substantial because EPIC-HDR is optimized by $d_\mathrm{ADISTS}$ rather than any of the evaluation metrics.

\begin{figure*}[t]
  \centering
    \subfloat[TMQI]{\includegraphics[height=0.44\linewidth]{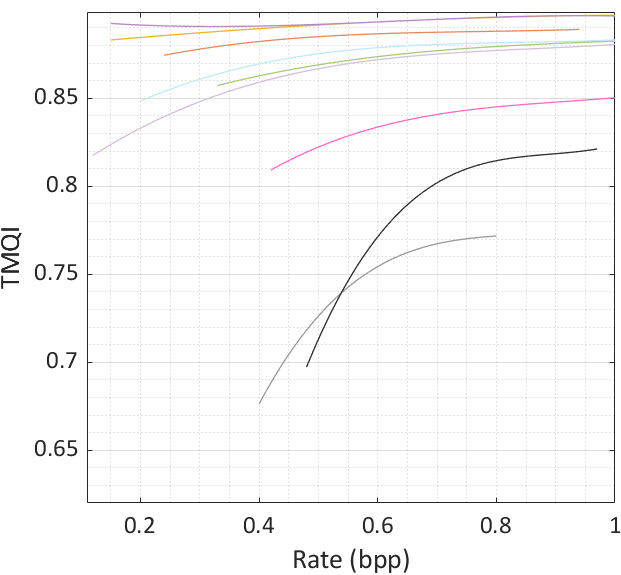}}
    \hfill
    \subfloat[NLPD]{\includegraphics[height=0.44\linewidth]{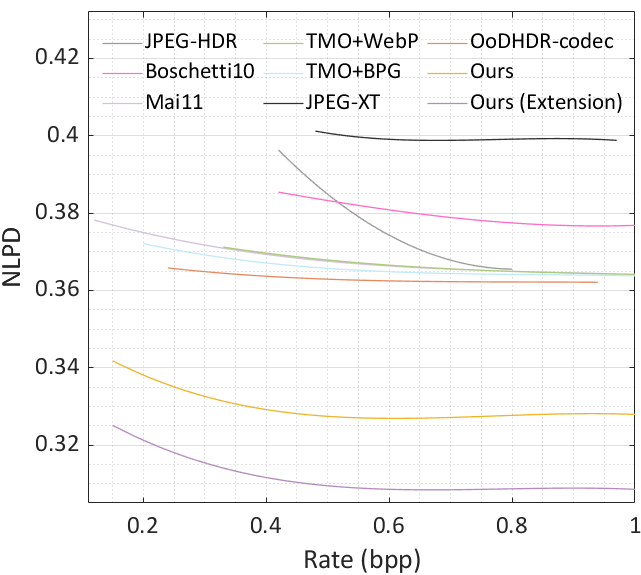}}
     \\
  \caption{LDR rate-distortion performance evaluation.} 
  \label{fig:ldr}
\end{figure*}

\noindent\textbf{LDR Image Comparison.} 
Fig.~\ref{fig:ldr} shows the average LDR rate-distortion curves, where EPIC-HDR and its extension produce better-quality LDR images in terms of both TMQI and NLPD at similar bit rates. 
Mai11 and its variants (TMO+BPG and TMO+WebP) achieve similar results due to the use of the same global TMO before compression. 
It is noteworthy that EPIC-HDR employs NLPD as the LDR image distortion function, resulting in greater improvements in this metric compared to TMQI.

\subsection{Qualitative Comparison}
\begin{figure*}[!t]
  \centering
\includegraphics[width=\linewidth]{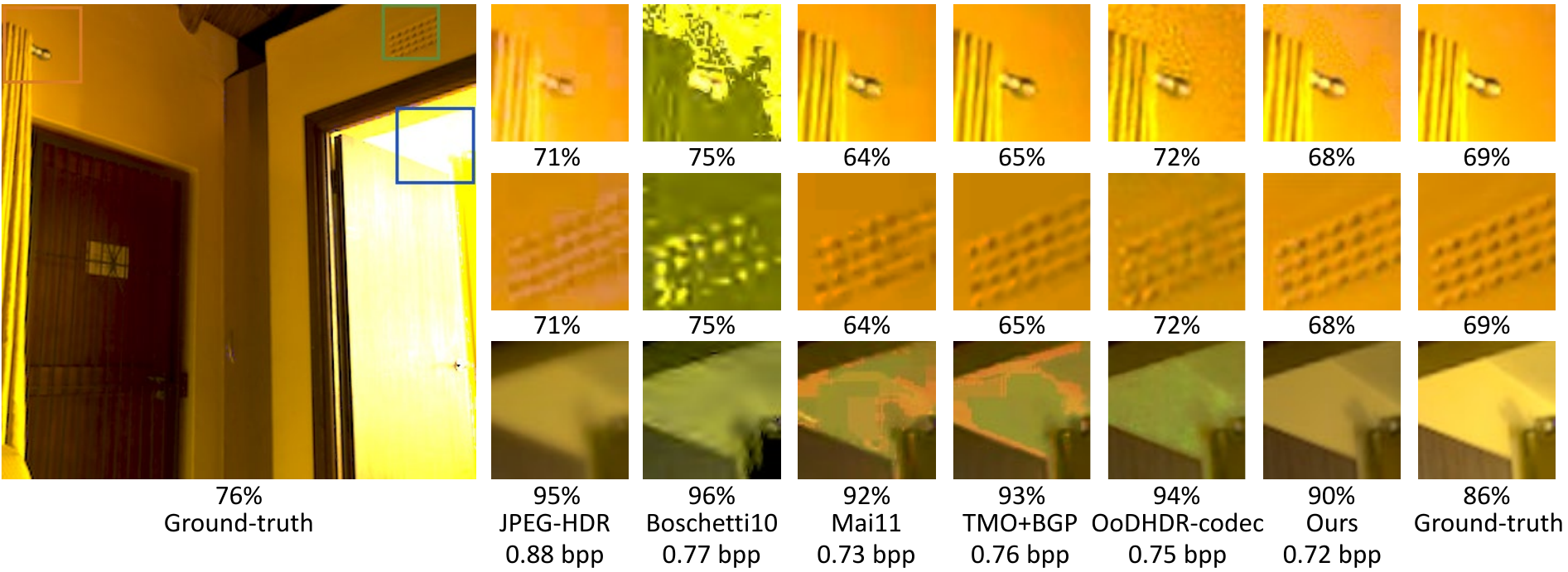}
  \caption{HDR image quality comparison on a ``Hallway'' scene. We set $e$ for the reference HDR image to be the $69$-th and $86$-th percentiles of the full dynamic range. The optimally matched $\hat{e}$ for each method (by Eq.~\eqref{eq:omeh}) is shown below.}
  \label{fig:cp-hdr2}
\end{figure*}

\begin{figure*}[!t]
  \centering
\includegraphics[width=\linewidth]{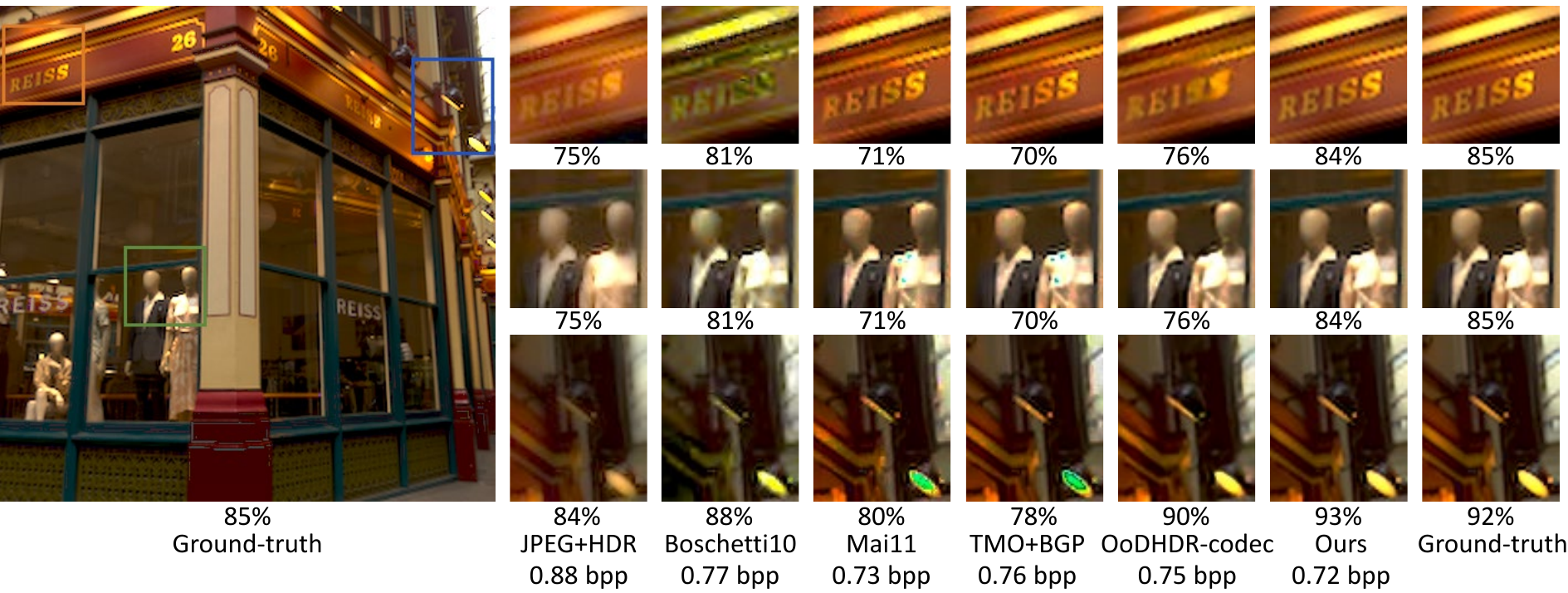}
  \caption{HDR image quality comparison on a  ``Storefront'' scene.}
  \label{fig:cp-hdr3}
\end{figure*}

\noindent\textbf{HDR Image Comparison.}
Figs.~\ref{fig:cp-hdr2} and~\ref{fig:cp-hdr3} compare the output HDR images at similar bit rates. 
JPEG-HDR exhibits blocking and blurring artifacts. 
Boschetti10 suffers from texture loss and color cast, while Mai11 and its variants sometimes exhibit blocky color distortions.
 OoDHDR-codec offers an improved visual appearance but falls short in rendering structural and textural 
details, especially in the text regions. 
In contrast, EPIC-HDR produces the overall highest-quality image, closest to the reference. 

\begin{figure*}[!t]
  \centering
  \includegraphics[width=\linewidth]{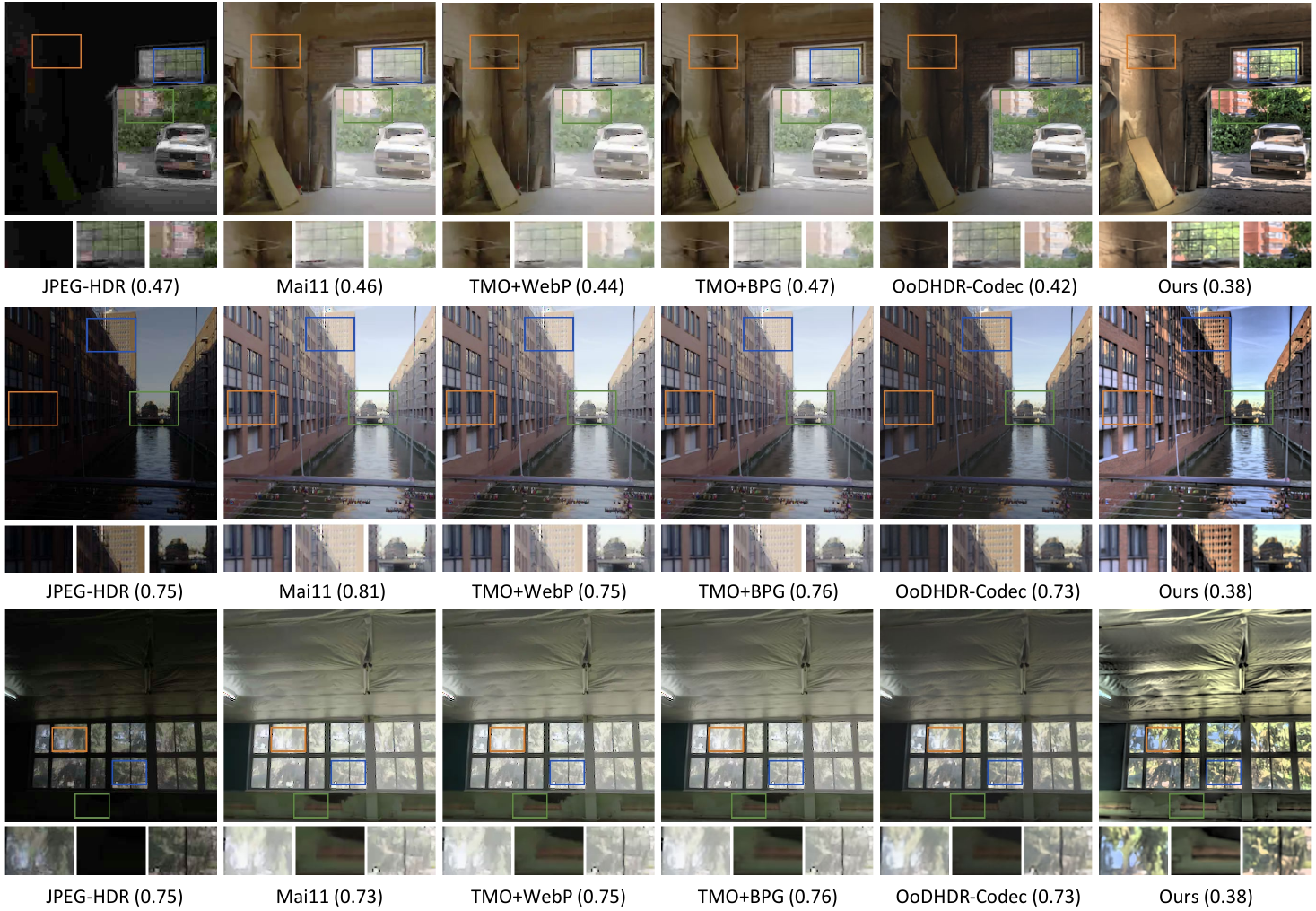}
  \caption{LDR image quality comparison on the ``Garage'', ``Watertown'' and, ``Indoor'' scenes, respectively, in which EPIC-HDR operators at a much lower bit rate.}
  \label{fig:cp-ldr-2}
\end{figure*}

\noindent\textbf{LDR Image Comparison.} 
Fig.~\ref{fig:cp-ldr-2} compares the output LDR images, where EPIC-HDR operates at a much lower bit rate. 
JPEG-HDR~\cite{ward2006jpeg} finds difficulty in restoring under-exposed details, and meanwhile suffers from blocking artifacts.
Mai11~\cite{mai2011optimizing} and its variants produce less-detailed and color-saturated images in high-exposed areas.
OoDHDR-codec~\cite{cao2022oodhdr} retains more details in high-exposed regions but sacrifices more details in low-exposed regions compared to Mai11~\cite{mai2011optimizing}.
In contrast, EPIC-HDR consistently delivers the best-quality images, excelling in detail preservation and color fidelity.
Additionally, Fig.~\ref{fig:extldr} shows the qualitative comparison between the LDR images generated by EPIC-HDR with different maximum scene luminances $\{10^{4}, 10^{5}, 10^{6}, 10^{7}\}$ $\rm cd/m^{2}$ and its extension. Besides automating EPIC-HDR, the extended version successfully balances detail reconstruction and noise suppression.

\begin{figure*}[!t]
  \centering
    \subfloat[HDR-VDP-3]{\includegraphics[height=0.45\textwidth]{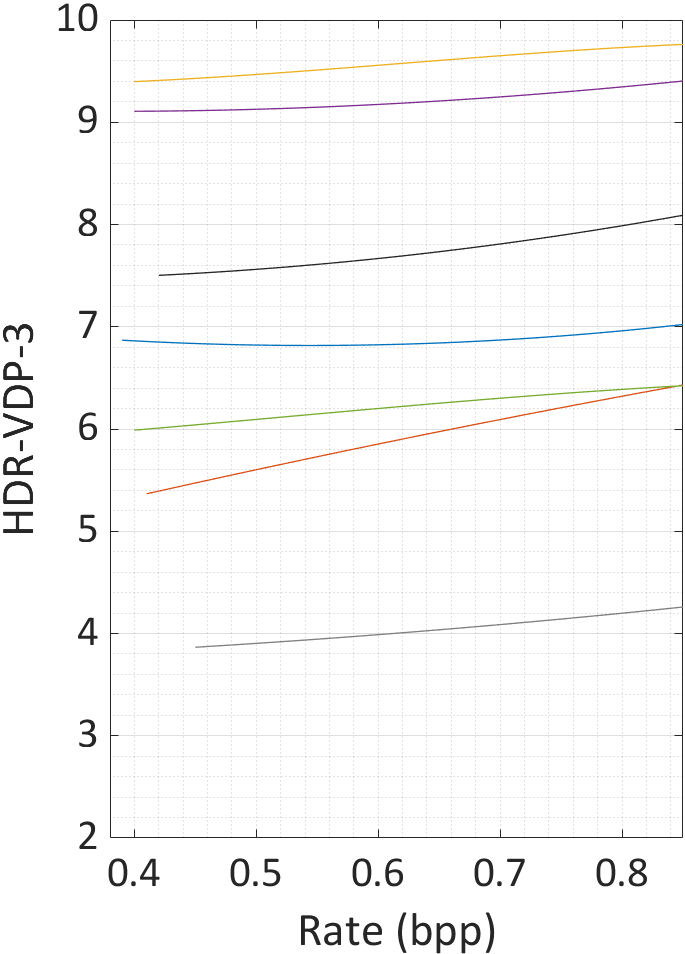}}
    \hfill
    \subfloat[PU21-CRF-PSNR]{\includegraphics[height=0.45\textwidth]{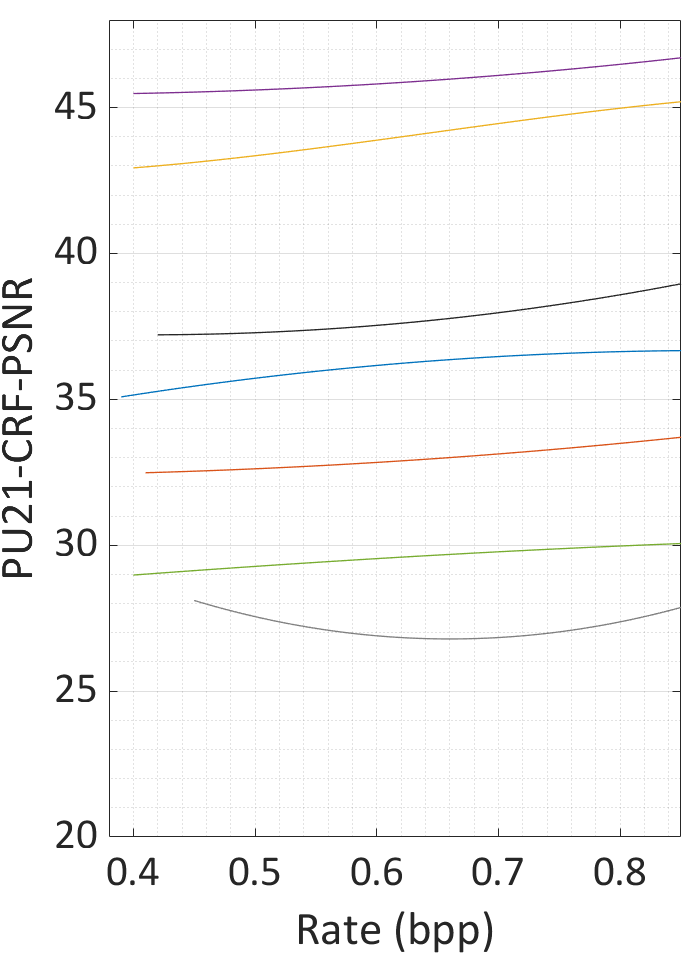}} 
    \hfill
    \subfloat[$d^\star_\mathrm{PSNR}$]{\includegraphics[height=0.45\textwidth]{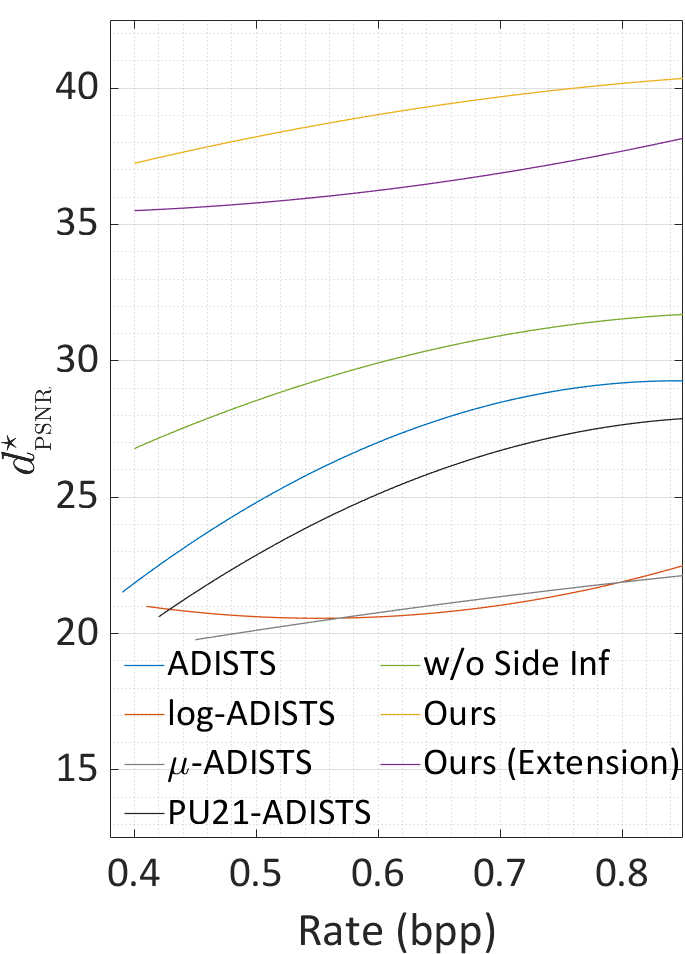}} 
  \caption{HDR rate-distortion curves of the EPIC-HDR variants.}
  \label{fig:abl}
\end{figure*}

\begin{figure*}[t]
  \centering
  \includegraphics[width=0.98\linewidth]{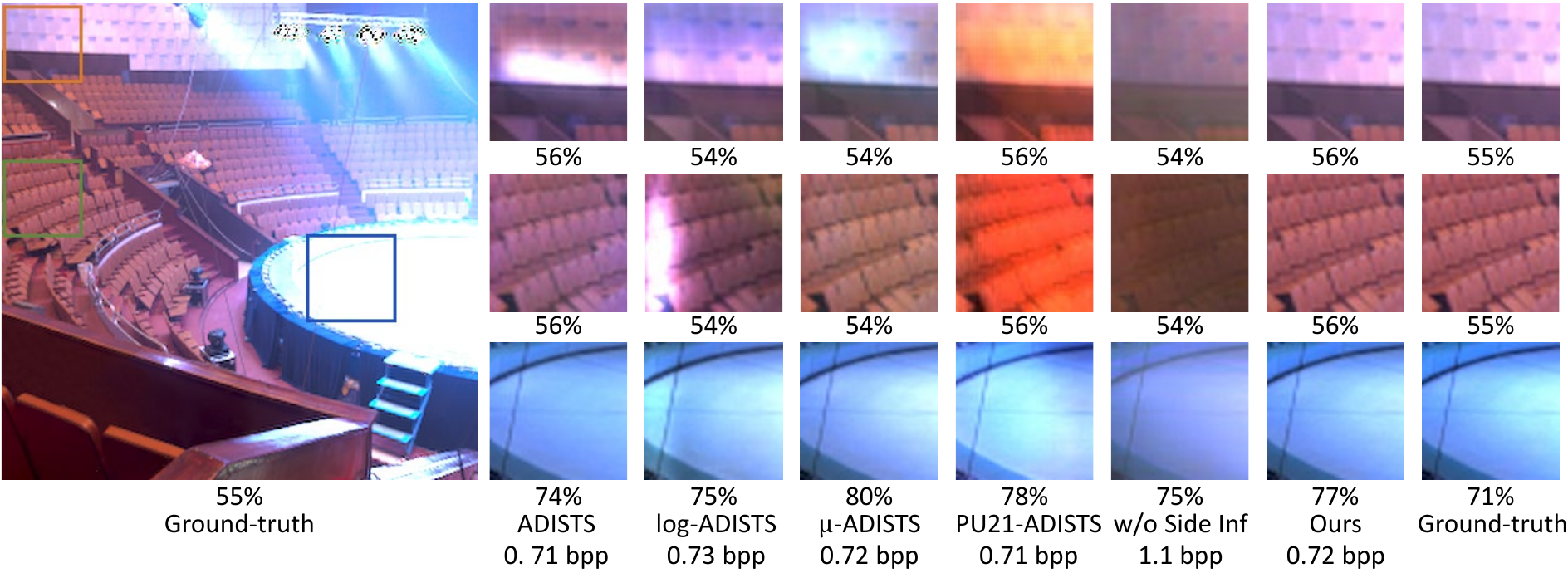}
  \caption{ HDR image quality comparison of different EPIC-HDR variants.}
  \label{fig:abl_vis}
\end{figure*}

\subsection{Ablation Experiments}
\noindent\textbf{HDR Image Distortion Function.}
We compare five different ``HDR'' image distortion functions: 1) ADISTS~\cite{ding2021locally}, 2) $\log$ encoded ADISTS ($\log$-ADISTS), 3) ADISTS computed in the tone-mapped domain via  $\mu$-law ($\mu$-ADISTS), 4) PU21-ADISTS, and 5) $d_\textrm{ADISTS}$. 
Fig.~\ref{fig:abl} shows the average HDR rate-distortion curves, where we find that $d_\textrm{ADISTS}$ (\ie, ADISTS equipped with the simple inverse display model~\cite{Cao2024perceptual}) is noticeably better than other perceptual encoding methods or TMOs as front-end pre-processing (see also Fig.~\ref{fig:abl_vis}).

\noindent\textbf{Effect of HDR Side Information.}  
Fig.~\ref{fig:abl} also shows the rate-distortion gains by
including HDR side information, which consumes a very low bit rate (approximately ranging from $0.04$ to $0.05$ bpp). As shown in Fig.~\ref{fig:abl_vis}, HDR side information clearly helps improve the overall color appearance and sharpness of the reconstructed HDR image.

\begin{figure*}[!t]
  \centering
   \subfloat[TMQI]{\includegraphics[height=0.43\textwidth]{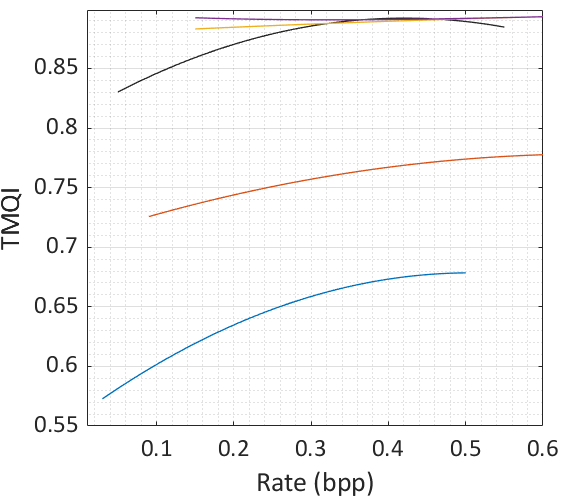}}   
   \hskip0em
   \subfloat[NLPD]{\includegraphics[height=0.43\textwidth]{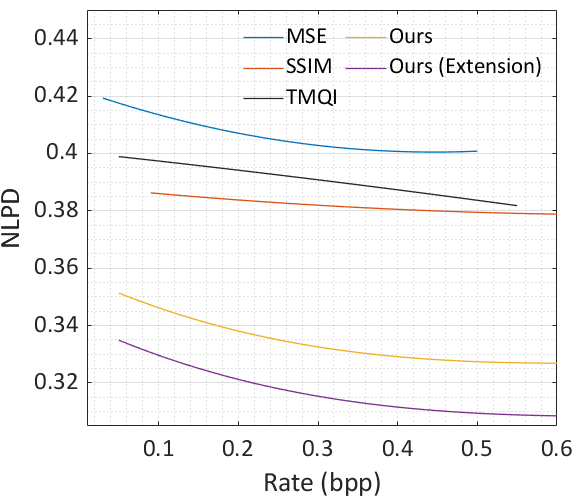}}
  \caption{LDR rate-distortion curves of the EPIC-HDR variants.} 
  \label{fig:abl2}
\end{figure*}

\noindent\textbf{LDR Image  Distortion Function.}
We switch NLPD~\cite{laparra2017perceptually} to three alternative LDR distortion functions: MSE, SSIM, and TMQI. 
Fig.~\ref{fig:abl2} illustrates the average LDR rate-distortion curves. EPIC-HDR and its extension generate better-quality LDR images in terms of both TMQI and NLPD. The TMQI-optimized variant performs favorably under TMQI as expected, but poorly in terms of NLPD.

\section{Conclusion and Discussion}
We have presented an end-to-end optimized HDR image compression system for perceptually optimal storage and display. Following the classic transform coding paradigm, EPIC-HDR transforms an HDR image into two sets of latent codes, which are subsequently quantized and losslessly compressed into two bitstreams. The first bitstream is used to generate an LDR image conditioned on the maximum scene luminance, ensuring backward compatibility with LDR displays. The second bitstream records HDR side information that assists in reconstructing the HDR image from the generated LDR image.

EPIC-HDR prioritizes perceptual optimization by adopting two perceptually aligned image distortion measures, both with reference to the uncompressed HDR image. Additionally, we have provided an automated extension of EPIC-HDR via the help of multi-exposure image fusion. A  promising future direction involves joint HDR image calibration (\ie, estimation of the maximum scene luminance) and compression to pursue perceptual optimality (rather than physical plausibility).

\section*{Acknowledgements}
This work was supported in part by the National Natural Science Foundation of China (62071407), the
Hong Kong RGC Early Career Scheme (2121382), and the Hong Kong ITC Innovation and Technology Fund (9440379 and 9440390).

\bibliographystyle{splncs04}
\bibliography{main}
\end{sloppypar}
\end{document}